# Spatiotemporal dynamic of COVID-19 mortality in the city of São Paulo, Brazil: shifting the high risk from the best to the worst socio-economic conditions

Patricia Marques Moralejo Bermudi[1], Camila Lorenz[1], Breno Souza de Aguiar[2], Marcelo Antunes Failla[2], Ligia Vizeu Barrozo[3] and Francisco Chiaravalloti Neto[1]

1 Departamento de Epidemiologia, Faculdade de Saúde Pública da Universidade de São Paulo, São Paulo, SP, Brazil

2 Gerência de Geoprocessamento e Informações Socioambientais (GISA) da Coordenação de Epidemiologia e Informação (CEInfo) da Secretaria Municipal de Saúde de São Paulo, SP, Brazil

3 Departamento de Geografia, Faculdade de Filosofia, Letras e Ciências Humanas e Instituto de Estudos Avançados da Universidade de São Paulo, São Paulo, SP, Brazil

## ABSTRACT

Currently, Brazil has one of the fastest increasing COVID-19 epidemics in the world, that has caused at least 94 thousand confirmed deaths until now. The city of Sao Paulo is particularly vulnerable because it is the most populous in the country. Analyzing the spatiotemporal dynamics of COVID-19 is important to help the urgent need to integrate better actions to face the pandemic. Thus, this study aimed to analyze the COVID-19 mortality, from March to July 2020, considering the spatio-time architectures, the socio-economic context of the population, and using a fine granular level, in the most populous city in Brazil. For this, we conducted an ecological study, using secondary public data from the mortality information system. We describe mortality rates for each epidemiological week and the entire period by sex and age. We modelled the deaths using spatiotemporal and spatial architectures and Poisson probability distributions in a latent Gaussian Bayesian model approach. We obtained the relative risks for temporal and spatiotemporal trends and socio-economic conditions. To reduce possible sub notification, we considered the confirmed and suspected deaths. Our findings showed an apparent stabilization of the temporal trend, at the end of the period, but that may change in the future. Mortality rate increased with increasing age and was higher in men. The risk of death was greater in areas with the worst social conditions throughout the study period. However, this was not a uniform pattern over time, since we identified a shift from the high risk in the areas with best socio-economic conditions to the worst ones. Our study contributed by emphasizing the importance of geographic screening in areas with a higher risk of death, and, currently, worse socio-economic contexts, as a crucial aspect to reducing disease mortality and health inequities, through integrated public health actions.

## 1. Introduction

Deaths from COVID-19, caused by the novel coronavirus of the severe acute respiratory syndrome (SARS-CoV-2), are considered avoidable because some collective and individual measures can prevent infection and because appropriate health assistance greatly reduces death risks. Nevertheless, the risk of dying extremely varies according to some individual and geographic risk factors resulting in health inequity observed in several parts of the world (Maroko et al., 2020; Wang and Tang, 2020) since the beginning of the pandemic. In Brazil, COVID-19 was first reported in the city of São Paulo on February 25, 2020 (Souza et al., 2020). Until August 3, four months after the first reported death, the disease had already resulted in 2,733,677 confirmed cases and 94,104 deaths in the country (Brasil. Ministério da Saúde, 2020a). At the moment, the United States of America and Brazil are the epicentre of the disease.

Brazil is the fifth country in the world in territory and population, classified as an upper-middle-income economy (The World Bank, 2020). Almost 60% of the Brazilian population is concentrated in 6% of the big cities among which São Paulo is the largest. The Brazilian Unified Health System (SUS) guarantees healthcare for all citizens and thousands of foreigners residing or passing through the country (Santos, 2018). Despite this universal and whole care, geographic differences in mortality rates across areas have been observed in the national and intra-urban scales. Spatial heterogeneity in population features, such as age, underlying health, household densities, partial lack of sanitation, socio-economic status, contact networks and, mobility patterns (Yang et al., 2020) have emerged as potential propellants of the spatiotemporal spread of the disease.

Since the beginning of the pandemic, mapping the disease occurrence and spread has become a powerful tool to track and support measures to slow down the infection transmission (Kamel Boulos and Geraghty, 2020) around the world and at the local scale. Web-based Geographical Information Systems (GIS) have allowed near real-time monitoring using map-centric dashboards (Early Alert, 2020; Johns Hopkins University, 2020; World Health Organization, 2020). Despite advances in the use of technology to reduce the impact of the pandemic, little is known about the spatiotemporal patterns of mortality from COVID-19, especially in an intra-urban setting. Studying the spatiotemporal dynamic of deaths instead of cases may help better evaluate inequity. All the health disadvantages accumulated over decades of life due to any kind of deprivation increases the risk of dying from COVID-19. The lack of a robust spatiotemporal analysis undermines the comprehension of the mitigation strategies to potentialize the disease-control efforts. Thus, this study aims to unveil the spatiotemporal dynamic of COVID-19 mortality at a fine granular level in the city of São Paulo considering the socio-economic context of the population. This approach may shed light on the urgent need for solid evidence on health inequities during the COVID-19 outbreak.

## 2. Materials and Methods

2.1 Study Area and Data Acquisition

This ecological study, based on COVID-19 secondary mortality data, was delineated in the city of São Paulo, state of São Paulo, Brazil (Figure 1), where live an estimated population of 11,869,660 inhabitants whose distribution over the area produces a mean demographic density of 7,803 inhabitants/km², in 2020 (Fundação SEADE, 2020).

When using COVID-19 data, underreporting is always an issue that deserves attention, even for mortality. To minimize the effect of possible sub notification in deaths, we analyzed confirmed and suspected deaths from COVID-19. Thus, death data comprised confirmed and suspected deaths occurred between March 15 and June 13, 2020, extracted on June 18 from the Mortality Information System (SIM), Mortality Information Improvement Program (PRO-AIM) from the Epidemiology and Information Coordination (CEInfo) of the São Paulo Health

Secretariat (SMS-SP). Confirmed deaths due to coronavirus infection corresponded to the code B34.2 (coronavirus infection disease), according to the International Classification of Diseases Tenth Revision (ICD-10). Suspected deaths are coded as U04.9 (severe acute respiratory syndrome).

The places of residence of the COVID-19 deaths were geocoded by CEInfo/SMS-SP according to the residence place using its databases and Google Maps API geocoding script that uses public places as the base map. Resulting geocoded addresses were validated comparing the road or ZIP code where the record was allocated with the original ZIP code. Geocoded data were assigned to the 310 sample areas of the Brazilian Institute of Geography and Statistics (IBGE), for which demographic and socio-economic census data are available (IBGE, 2010). We considered these areas as the spatial units in our models (Figure 1).

The records, including the basic cause, age, sex, the date of the death according to the epidemiological week (EW) (Brasil. Ministério da Saúde, 2020b) and the sample areas of residence were obtained by formal request to the São Paulo Electronic Information System (e-SIC, protocol 48567), whose data information is hosted in an open session, on the municipality's transparency portal for public access (São Paulo. Prefeitura, 2020). Here, we named this information as e-SIC database. It was not necessary to submit this study to an Ethics Committee because we did not have access to the names and addresses. Thus, the use of secondary data, without personal identification and in the public domain, dispenses with the need for prior approval by the Ethics Committee on Research with Human Beings as per Resolution No. 510/2016, of the National Health Council (Guerriero, 2016).

We also used the data available in Tabnet-DataSUS and named this information as Tabnet database. Tabnet is an app, available in <https://www.prefeitura.sp.gov.br/cidade/secretarias/saude/tabnet/>, provided by the Municipal Health Department of São Paulo and developed by DataSUS. This app provides free access (to any user) to population databases and to the database information systems of SUS, such as the Mortality Information System (SIM), which is supplied by the Secretariat's Program (PRO-AIM). Through the Tabnet app, it is possible to perform tabulation and crossing several variables of interest such as epidemiological week, sex, age group, and specific cause. The databases are updated periodically.

As a measure of the socio-economic context of the population, since individual-level data are not available in the mortality database, we used a socio-economic index especially elaborated for health research. The GeoSES index (Barrozo et al., 2020) was developed using Principal Component Analysis, starting with 41 variables. The index conceives the socio-economic context preserving seven dimensions based on the theoretical background (Duncan et al., 2002; Krieger et al., 1997): education, mobility, poverty, wealth, income, segregation, and deprivation of resources and services. The index was defined in three scales: national, Federative Unit and, intra-municipal. Figure 1 presents GeoSES for the sample areas of the city of São Paulo, showing that the areas with the best socio-economic conditions (GeoSES equal to or close to 1) are located in the central part of the city and that these conditions deteriorate towards the periphery, where they reach the worst levels (GeoSES equal to or close to -1). It has been shown to be useful in studies of mortality from avoidable causes of deaths (from 5 to 74 years old) due to interventions at the Brazilian health system in the national scale and, mortality from circulatory system diseases in the city of São Paulo (Barrozo et al., 2020).

## 2.2 Data Analysis

We used the information of confirmed (B34.2) and suspected (U04.9) death from COVID-19 available for the entire city of São Paulo from EW 11th to 29th of Tabnet database to calculate the weekly mortality rates of confirmed (B34.2), suspected (U04.9) and total (B34.2 + U04.9) COVID-19 deaths. We did the same using the e-SIC database from EW 11th to 24th. We exclude from the information e-SIC database the COVID-19 deaths occurred on 25th because the data from this week was incomplete (it was extracted on 18th June 2020 and included only part of EW 25th). These rates were obtained dividing the respective numbers of deaths in each week by the total population of the city and presented as death per 100,000 inhabitants-week. In the sequence, we obtained the mortality rates for confirmed, suspected, and total COVID-19 deaths by sex and age for the entire period from EW 11th to 24th using Tabnet and e-SIC databases. These comparisons between the data of these two sources were useful to evaluate how complete was the data we used for the spatial and spatiotemporal analysis. For the calculation of the mortality rates by sex and age, we excluded the data without this information.

As only one suspected COVID-19 death occurred in EW 11th, we restricted our spatial and spatiotemporal analysis from EW 12th to 24th and spatial or spatiotemporal architecture was considered in all the models we performed. We first modelled the confirmed and total COVID-19 deaths using spatiotemporal models only with the intercept and random effects accounting for spatial and temporal autocorrelation and the interaction between them. The spatial dependence was modelled considering a Besag-York-Mollié (BYM) model with two components representing the spatially structured and non-structured random effects (Blangiardo et al., 2013; Blangiardo and Cameletti, 2015). These two components were considered independent one for another and followed the parametrization proposed by Simpson et al. (2017). The temporal dependence was modelled by a non-structured random effect and a structured random effect given by a random walk autoregressive model of first-order (RW1). The interaction between space and time was modelled considering spatial and temporal non-structured random effects (Blangiardo and Cameletti, 2015).

The number of confirmed and total COVID-19 deaths by EW and sample areas were modelled using Poisson and zero-inflated Poisson probability distributions in a latent Gaussian Bayesian model approach. We considered the expected confirmed and total COVID-19 death for each EW and spatial unit as an off-set in these models. The expected deaths were estimated by indirect standardization taking into account the age and sex structure of each sample area and the mortality rates for the entire study period and city. This, therefore, enables us to interpret the outcomes of our analysis as relative risks (RR) concerning the mortality rates for the entire period and city. We obtained, from these models, the temporal and spatiotemporal RR. Subsequently, we introduced the socio-economic covariate (GeoSES) in these models and obtained the corresponding RR.

Finally, we used a spatial approach to model the confirmed and total COVID-19 death by EW to evaluate the role of the socio-economic covariate in each one of the EW. For doing this, we considered spatial models with intercept, BYM spatial random effects, and the GeoSES as a covariate. The expected COVID-19 deaths were obtained in a similar way for the spatiotemporal models, but considering the entire city mortality rates for each EW, allowing us to interpret the RR concerning the entire city mortality rates for each EW.

We did our models in a Bayesian context using the integrated nested Laplace approximation (INLA) approach (Rue et al., 2009). We selected our best models using the Deviance Information Criterion (DIC) so that the best-adjusted models were those with lower DIC values (Blangiardo and Cameletti, 2015). We used non-informative priors for the fixed effects and priors with penalized complexity for the precision parameters of the random effects (Simpson et al., 2017). We ran our models in the R environment (R Core Team, 2019).

## 3. Results

We found 14,753 confirmed and suspected COVID-19 deaths in the Tabnet database, from EW 11th to 29th, and 10,760 in the e-SIC database, from EW 11th to 25th, in the city of São Paulo. We removed 67 deaths from the e-SIC database because they were referent to the EW 25th which was not completed when the data were extracted. Figure 2 shows the mortality rates for COVID-19 for both sources of data, considering the confirmed, suspected and total deaths. The curves from the e-SIC database are similar to the curves with data from the Tabnet database and the differences among these curves in EW 23th and 24th are related to a delay in the notification of the COVID-19 deaths.

Table 1 shows the numbers and mortality rates of confirmed, suspected, and total COVID-19 deaths obtained from e-SIC and Tabnet databases from EW 11th to 24th by sex and age. To build Table 1, we excluded six deaths with ignored age and three deaths with ignored sex from the 10,693 (10,760 – 67) death in the e-SIC database and we excluded six ignored age and four deaths with ignored sex from the 11,098 deaths in the Tabnet database. We can observe that mortality was higher for males and that it increases as age increases, corresponding to 460.9 deaths per 100,000 inhabitants (in fourteen weeks) for people aged 60 years or older. This pattern of increased mortality as age increases is also observed when making a greater stratification among people aged or older, as shown in supplementary material 1.

Table 1 shows that the data we used to build our spatial and spatial models (e-SIC database) is close to the municipality's official data on the pandemic (Tabnet-DataSUS). For doing these models we excluded, from the e-SIC database, a suspected COVID-19 death occurred on EW 11th (the first one in the city) and nine with ignored sex or age and 68 whose addresses were not geocodified and did not have the sample area codes. We achieved a high geocoding success rate of 99.3%, once 10,619 records were geocoded using address data out of a total of 10,692 (excluding the first death in EW 11th). Of these, 5,837 refer to ICD B34.2 (99.4% of the initial total of 5,875 records) and 4,782 to ICD U04.9 (99.2% of the total of 4,817 records).

During this period, 30.6 and 17.4% of the sample areas had zero confirmed and total COVID-19 deaths and, therefore. The DIC values of the spatiotemporal models with Poisson probability distribution were lower than the value of zero-inflated Poisson distribution (Supplementary material 2). Considering each week separately, the amount of zeros deaths in the sample areas varied from 6.1 to 85.5%. The DIC values for the spatial models with Poisson probability distribution were, in most cases, lower than the values with zero-inflated Poisson distribution, and, in the cases that this did not occur, they are very close to each other

(Supplementary material 3). From these results, we considered the best-adjusted spatial and spatiotemporal models that ones with the Poisson distribution.

First, we present the results of the models with spatiotemporal architecture only with intercept. We have, in Figure 3, the temporal RR from EW 12th to 24th, where we can see that maximum RR occurred on EW 20th for the total COVID-19 deaths and in EW 23 for the confirmed ones in our study period. These results, even adjusted for the temporal autocorrelation, are similar to those presented in Figure 2. Considering the data presented in Figures 2 and 3, the apparent pattern of the temporal curves shows a tendency to stabilize, but a new rise in mortality cannot be ruled out soon.

Figures 4 and 5 show the posterior means of the spatiotemporal RR for the sample areas and EW, respectively, for confirmed and total COVID-19 deaths. Apart from the fact that the RR is greater for the total deaths than the confirmed ones, the distribution of the RR is similar between them, following the behaviour of the temporal RR. In the first two EW, the sample areas presented lower values of RR that increased over time. However, this increase occurred with greater intensity in the peripheral areas.

Next, we performed the spatiotemporal models, now taking into account the socio-economic variable (GeoSES). Table 2 shows the spatiotemporal RR and the 95% credibility intervals for GeoSES obtained for the models with confirmed and total COVID-19 deaths. In both models, it is noted that the high socio-economic level was shown to protect against the risk of dying from COVID-19 throughout the study period. Thus, the increase of one unit in the socioeconomic indicator represented a 25% reduction in the risk of dying from COVID-19, for the model using confirmed deaths, and a 33% reduction in the risk of dying, for the model using COVID-19 total deaths. Moreover, the risk of dying from COVID-19 in the sample areas with the best socio-economic conditions (GeoSES close to 1), about the areas with the worst ones (GeoSES close to -1) was 50% lower for the model with confirmed deaths and 66% lower for the model with the total deaths.

Finally, we performed the spatial modelling of confirmed and total COVID-19 deaths in the EW separately, taking into account the socio-economic covariate. Figure 6 shows RR and 95% credibility intervals for GeoSES, according to the confirmed and total COVID-19 deaths for each one of the EW. We identified a shift in the pattern of the relationship between COVID-19 mortality and socio-economic status over time. The best socio-economic level presented itself as a risk factor for COVID-19 deaths in the first two EW in the city of São Paulo. From EW 15th, for the total death, and from EW 16th, for confirmed and total death, the worst socio-economic level became a risk factor. Even if some values were not significant, there was a continuous decrease in RR from EW 12th to 17th, followed by stabilization.

## 4. Discussion

This is the first population-based study on the evolution of the spatiotemporal pattern of mortality from COVID-19 in the intra-urban setting of the largest city of Brazil. Using two different datasets, analyzing confirmed and confirmed + suspected deaths separately allows evaluating how uncertainty would impact the association between RR and the socio-economic context. The robust models by EW clearly show when the high risk of death shifted from the best to the worst socio-economic conditions in the city.

Our findings showed that the most critical period regarding mortality by COVID-19 in the city of Sao Paulo occurred between EW 20th and 23th, followed by an apparent stabilization of the temporal trend. However, it is not possible to predict a future scenario, as social distancing measures have been relaxed in the city since the 25th week (MS, 2020) this could increase the number of infected people and, consequently, the number of deaths. Albeit social distancing alone seems not to be enough to contain COVID-19, many studies frequently concluded that it is a critical component of outbreak control (Nussbaumer-Streit et al., 2020). It is important to point out that both total deaths and only confirmed deaths showed similar patterns in our study, despite its differences. The suspected deaths need, for the one hand, to be treated with caution, because they may be not COVID-19 and, for the other hand, to consider them is one of the strengths of the study. First of all, there is a delay in the confirmation of the suspected cases, consequently part of the suspected deaths will be confirmed as COVID-19 deaths in the future. Furthermore, part of them would be confirmed, if there were no difficulties related to the strict case definition, which requires testing often not available or not performed in the appropriate time window. From this point of view, the amplitudes of variation in rates and relative risks obtained from confirmed and total deaths could be considered as lower and upper limits (or vice versa) for the magnitudes of these measures.

  The elderly population represents one of the groups that are more prone to the infection and symptomatology by COVID-19 in the city of Sao Paulo, with a higher risk of death for men after the seventh decade of life, similar to statistics found for China and the United States (CDC. The Novel Coronavirus Pneumonia Emergency Response Epidemiology Team, 2020). Recently Souza et al (2020) analyzed the Brazilian population and also found that most COVID-19 deaths were male, and the most frequent comorbidities were cardiovascular disease and diabetes. Behavioural elements, especially posture that may prejudice adherence to lockdown measures, have been demonstrated as potentially crucial in determining susceptibility to SARS-CoV-2 (Pawlowski et al., 2008; Raisi-Estabragh et al., 2020). This unequal death ratio in men may be interpreted considering a lot of factors: their comparatively higher presence of comorbidities (i.e., hypertension, diabetes, cardiovascular disease, and chronic lung disease) (Sharma et al., 2020), higher risk behaviours (i.e., smoking and alcohol use), occupational exposure (Global Health 5050, 2020) and, sex differences in immune responses (Klein and Flanagan, 2016). In contrast, there may be other social and behavioural characteristics that favour women, with previous studies proposing women are more likely than men to adopt hand hygiene practices (Johnson et al., 2003) and seek preventive care (Bertakis et al., 2000).

  The spatial distribution of suspected and confirmed deaths by COVID-19 in the city of Sao Paulo shows inequalities, with spatial dependence and positive correlation associated with socio-economic factors of the areas, remarkably similar to the results of Maciel et al. (2020). Our findings reveal that socio-economic status acts as a protective factor against the risk of dying from COVID-19. In the models with only confirmed deaths and with all deaths, the increase of one unit in the socio-economic indicator represented, respectively, a 25% and 33% decrease in the risk of dying. The first observation is that, when considering all deaths, the protective effect of the socio-economic level is more evident, showing that there must be a higher incidence of suspected deaths in the less-favoured areas concerning the most favoured areas (in areas with better socio-economic level, access to confirmation for COVID-19 probably is more available). A study conducted by Souza et al. (2020) reinforces this finding. They compared the spatial pattern of confirmed cases of COVID-19 and severe acute respiratory infection with unknown aetiology with the socio-economic level in the metropolitan region of São Paulo and found that the firsts were more associated with better levels and the seconds. They pointed out the degree of underreporting of COVID-19 cases should increase with a decrease in socio-economic status. Therewith, our results have confirmed the association between COVID-19 and human development, pointing to the importance of geographic screening in locations with a potential for local infectious transmission as a fundamental aspect to coordinate better actions against the pandemic (Maciel et al., 2020).

The low levels of socio-economic position reveal that not only the vulnerability of the population but also the difficulties in health services concerning diagnosis and treatment of the condition, similar to the overview of fragility expected from health services in Brazil (Ribas et al., 2020) and in the countries of Latin America to face the pandemic (Rodriguez-Morales et al., 2020). Besides, living conditions also may be strongly influenced by the low income in different ways, such as residence in more poor neighbourhoods and housing conditions, particularly confined or overcrowded housing (Khalatbari-Soltani et al., 2020), which has been related with a greater risk of contagion from several other pathogens, such as *Helicobacter pylori* (Webb et al., 1994), tuberculosis (Gupta et al., 2004) or Epstein–Barr virus (Gares et al., 2017). Besides, a person's employment may expose them to different risks related to the type of job (Khalatbari-Soltani et al., 2020). Work requiring continuous human contact, such as caring for people or interaction with others means that risk of infection dissemination through droplets of aerosol is higher (Rule et al., 2018). Regarding COVID-19, studies showed that occupation is an explicit determinant of contagion and a secondary determinant of COVID-19 severity and deaths by the association between occupational social class and comorbidities (Khalatbari-Soltani et al., 2020). For example, workers such as cleaners, retail staff, teachers, or healthcare workers suffer the direct impact of the COVID-19 incidence (Koh, 2020). People with underprivileged socio-economic conditions are more prone to be exposed to job stress including burnout syndrome and unemployment, which may contribute to disrupted immune and inflammatory system responses (Berger et al., 2019; Nakata, 2012) as well as a higher risk for comorbidities for COVID-19 (Kivimäki and Kawachi, 2015). Until now, both debilitated immunity and the existence of comorbidities are recognized risk factors of COVID-19 severity (Khalatbari-Soltani et al., 2020).

We showed that the first cases of deaths occurred in the neighbourhoods with the best socio-economic position in the city of Sao Paulo. This may be related to the fact that all of the infected subjects had been abroad (MS, 2020). In the first two weeks, the best socio-economic level was presented as a risk factor. Then there was a change in the spatial pattern: from the fourth week onwards, the worst socio-economic level becomes a risk factor. Similarly, Souza et al. (2020) showed a higher risk of diagnosed COVID-19 cases in census tracts with higher per-capita income in the Sao Paulo metropolitan region during the early phase of the COVID-19 epidemic. After these first cases in richer areas, the virus started to circulate in the suburbs of the city, with high population density and worsened sanitary conditions (Silva and Muniz, 2020), and probably explain its fast transmission. The city of Sao Paulo is particularly vulnerable because it is the most populous in the country, with approximately 12 million inhabitants (IBGE, 2020), and is highly connected within Brazil and around the world. Its main airport, the São Paulo-Guarulhos International Airport, is the largest in Brazil, with non-stop passenger flights to 103 destinations in 30 countries (Rodriguez-Morales et al., 2020).

In this study, we used the SIM database instead of SIVEP-Gripe, unlike other studies (Souza et al., 2020). The recommendation of the Ministry of Health to register the notification of death and the monitoring of mortality from this system, in practice, leads to a longer time between the event and the use of information by the teams since the registration of death occurs initially in the SIM, based on the death certificate and requires health service teams and/or COVISA, who access the systems, to insert the evolution for each notified case or the notification of death for cases not previously notified. These characteristics, combined with the coping strategy adopted by SMS-SP, which did not include mass testing, motivated the option of using SIM data considering the confirmed and suspected diagnoses for analyzing mortality caused by coronavirus infection in the analyzes performed.

Our study findings must be considered in the context of several assumptions and data limitations. We associate patient's addresses or postcodes to area-based socio-economic positions using the geolocalisation, which may provide some insight into the likelihood of exposure to health factors and COVID-19 risks. This approach is frequently used as representative for the individual socio-economic condition; nevertheless, they are not a perfect picture of individual circumstances, could underestimate the magnitude of social disproportion related to individual social measures (Lamy et al., 2019) and are best employed in complement with individual-level variables to reflect geographical or aggregate-level risks (Khalatbari-Soltani et al., 2020). We highlight that our spatial analysis is subject to methodological limitations caused by ecological fallacy and the modifiable areal unit problem. These constraints are intrinsic to any spatial analysis using aggregated data (Subramanian et al., 2009). Even so, our study contributes to healthcare planning measures and for future precision studies focusing on the effects of social health factors on COVID-19 deaths. Also, one of the strengths of our study was dealing with deaths by COVID-19 instead of using the cases, due to the better accuracy and reliability of the data. When we consider only the cases, there may be many underreporting asymptomatic patients that can hamper the conclusions.

## 5. Conclusions

We used models with spatial and spatio-temporal architectures to investigate the spatial and spatio-temporal patterns of confirmed and total (confirmed and suspected) COVID-19 deaths in the city of São Paulo. The obtained results from considering both categories showed differences regarding the magnitude of rates and RR, however, there were no differences concerning the conclusions we achieved. The maximum risk of dying from COVID-19 occurred between EW 20th and 23th, followed by an apparent stabilization of the temporal trend, but we did not rule out a new rise in mortality soon. We identified that the high socio-economic level was shown to protect against the risk of dying from COVID-19 throughout the study period. However, this was not a uniform pattern, since we identified a shift in the risk of dying from COVID-19 in the city of São Paulo over time: from high risk in the best socio-economic contexts in the first two EW to high risk in the worst contexts, from EW 16th ahead. Concerning sex and age, men and elderly people presented the highest risk from dying of COVID-19. Our study has corroborated the relationship between COVID-19 mortality and socio-economic condition, revealing the importance of geographic screening in areas with a higher risk for deaths as a crucial aspect to integrate better actions to face the pandemic.

**Funding**

This work was supported by the Conselho Nacional de Desenvolvimento Científico e Tecnológico [grant numbers 301550/2017-4 to LVB and 306025/2019-1 to FCN]; and the São Paulo Research Foundation (FAPESP) [grant number 2017/10297-1 to CL].

**Figures and Tables**

**Table 1.** Number and mortality rates (per 100,000 inhabitants in fourteen weeks) of suspected (U04.9), confirmed (B34.2), and total (U04.9 + B34.2) COVID-19 deaths, according to e-SIC and Tabnet databases, sex, and age. City of São Paulo, 11th to 24th epidemiological weeks, 2020. Data source: Deaths: Sistema de Informações sobre Mortalidade – SIM/PRO-AIM – CEInfo –SMS-SP. Population: Fundação SEADE. Tabnet database was updated on 7/23/2020 and e-SIC database was provided on 6/08/2020.

|  | e-SIC database (provided on 6/18/2020) | | Tabnet database (updated on 7/23/2020) | | Ratio: Tabnet database / e-SIC database |
|---|---|---|---|---|---|
|  | Nº of deaths | Mortality rate (per 100,000 inhab.) | Nº of deaths | Mortality rate (per 100,000 inhab.) |  |
| **Male sex** | | | | | |
| **Confirmed** | 3350 | 59.3 | 3861 | 68.3 | 1.2 |
| **Suspect** | 2507 | 44.4 | 2228 | 39.4 | 0.9 |
| **Total** | 5857 | 103.6 | 6089 | 107.7 | 1.0 |
| **Female sex** | | | | | |
| **Confirmed** | 2524 | 40.6 | 2906 | 46.7 | 1.2 |
| **Suspect** | 2309 | 37.1 | 2099 | 33.8 | 0.9 |
| **Total** | 4833 | 77.7 | 5005 | 80.5 | 1.0 |

|  | Total | | | | |
|---|---|---|---|---|---|
| Confirmed | 5875 | 49.5 | 6768 | 57.0 | 1.2 |
| Suspect | 4818 | 40.6 | 4330 | 36.5 | 0.9 |
| Total | 10693 | 90.1 | 11098 | 93.5 | 1.0 |
|  | Total: 0 to 19 years old | | | | |
| Confirmed | 17 | 0.6 | 20 | 0.7 | 1.2 |
| Suspect | 41 | 1.4 | 41 | 1.4 | 1.0 |
| Total | 58 | 1.9 | 61 | 2.0 | 1.1 |
|  | Total: 20 to 39 years old | | | | |
| Confirmed | 236 | 6.2 | 284 | 7.5 | 1.2 |
| Suspect | 208 | 5.5 | 172 | 4.5 | 0.8 |
| Total | 444 | 11.7 | 456 | 12.0 | 1.0 |
|  | Total: 40 to 59 years old | | | | |
| Confirmed | 1145 | 35.5 | 1324 | 41.1 | 1.2 |
| Suspect | 825 | 25.6 | 706 | 21.9 | 0.9 |
| Total | 1970 | 61.2 | 2030 | 63.0 | 1.0 |
|  | Total: 60 years old or older | | | | |
| Confirmed | 4475 | 241.5 | 5140 | 277.3 | 1.1 |
| Suspect | 3739 | 201.7 | 3405 | 183.7 | 0.9 |
| **Total** | 8214 | 443.2 | 8545 | 461.1 | 1.0 |

**Table 2.** Posterior means of the relative risks (RR) and 95% credibility intervals for the socio-economic covariate (GeoSES) obtained with the spatiotemporal models for confirmed and total COVID-19 deaths. City of São Paulo, 12th to 24th Epidemiology Weeks, 2020.

| COVID-19 deaths | Covariate | RR posterior means | 95% credible interval | |
|---|---|---|---|---|
| | | | 0.025 quantil | 0.975 quantil |
| Confirmed | Intercept | 0.82 | 0.78 | 0.85 |
| | GeoSES | 0.75 | 0.69 | 0.82 |
| Total | Intercept | 0.74 | 0.71 | 0.77 |
| | GeoSES | 0.67 | 0.62 | 0.72 |

**Figure 1.** a) South America, Brazil, State of São Paulo, city of São Paulo. b) Distribution of socio-economic Index of the Geographic Context for Health Studies (GeoSES) according to the sample area and delimited by administrative districts (DA), city of São Paulo, 2010.

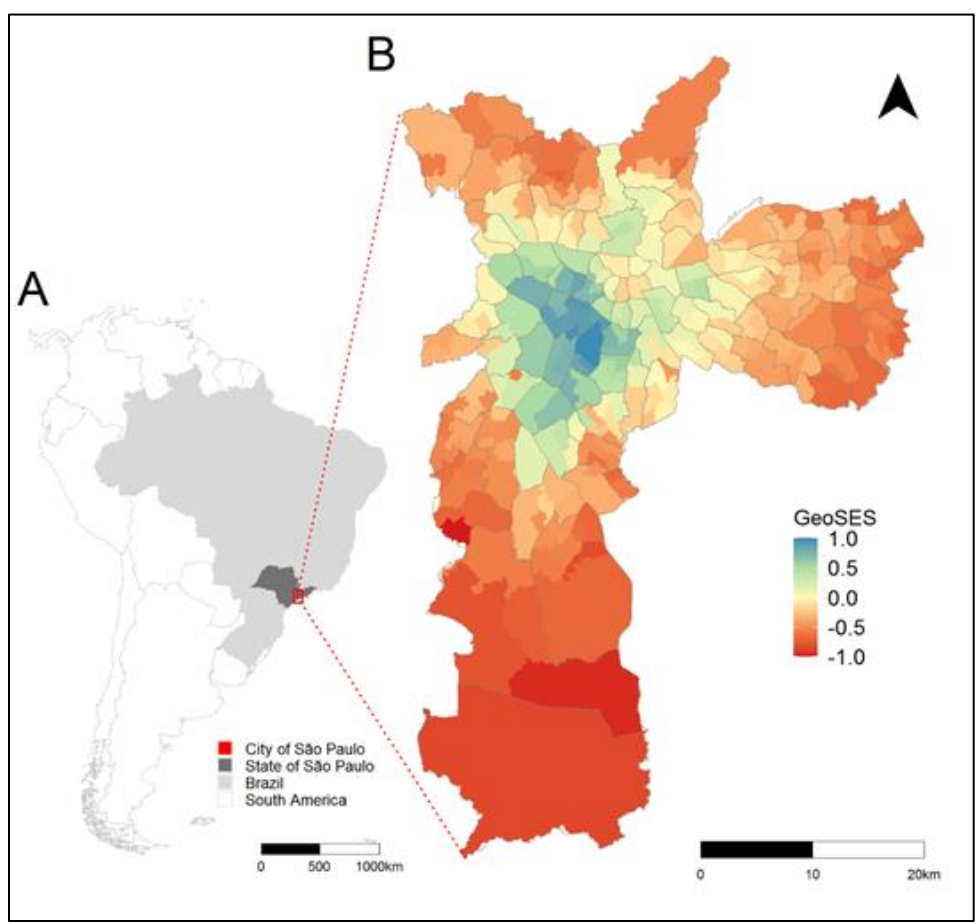

**Figure 2.** Distribution of mortality rates (per 100,000 inhabitants-week) of suspected (U04.9), confirmed (B34.2), and total (U04.9 + B34.2) COVID-19 deaths, according to e-SIC and Tabnet databases and epidemiological week. City of São Paulo, 2020. Data source: Deaths: Sistema de Informações sobre Mortalidade – SIM/PRO-AIM – CEInfo –SMS-SP. Population: Fundação SEADE.  Tabnet database was updated on 7/23/2020 and e-SIC database was provided on 6/08/2020.

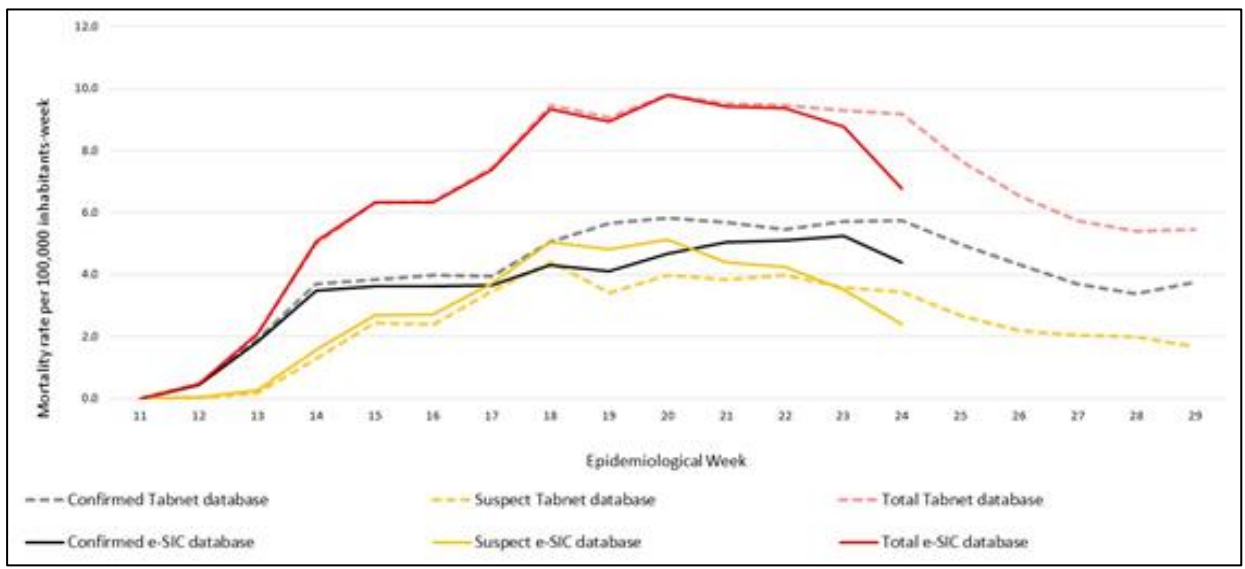

**Figure 3.** Posterior means of the temporal relative risks (RR) of COVID 19. City of São Paulo, 12th to 24th Epidemiological Week, 2020.

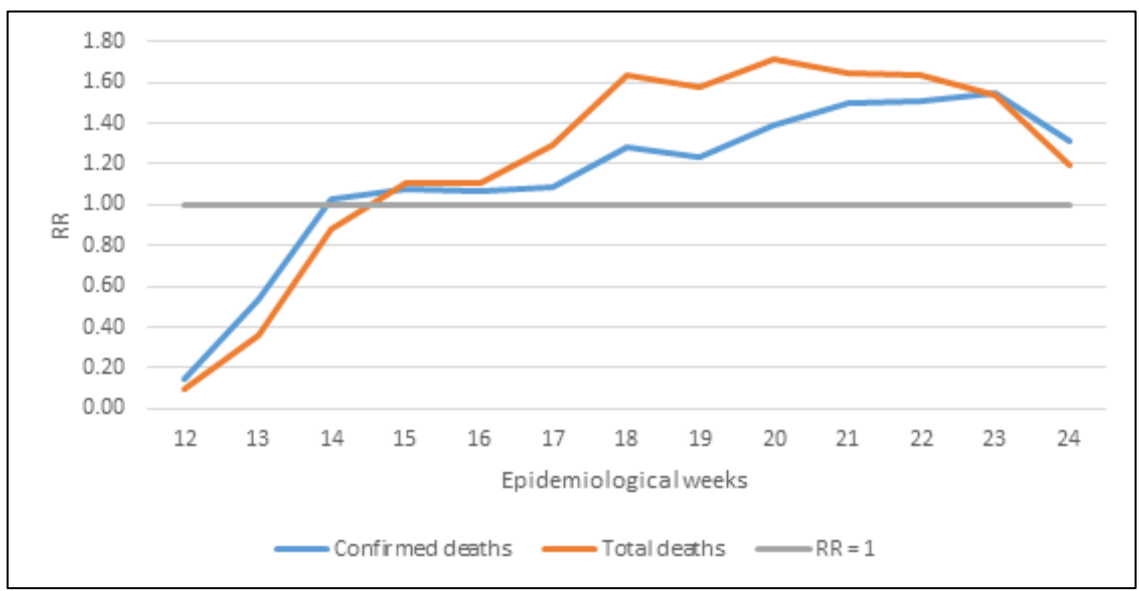

**Figure 4.** Posterior means of the spatiotemporal relative risks (RR) for confirmed COVID-19 deaths. Sample areas of the city of São Paulo, 12th to 24th Epidemiological Week, 2020.

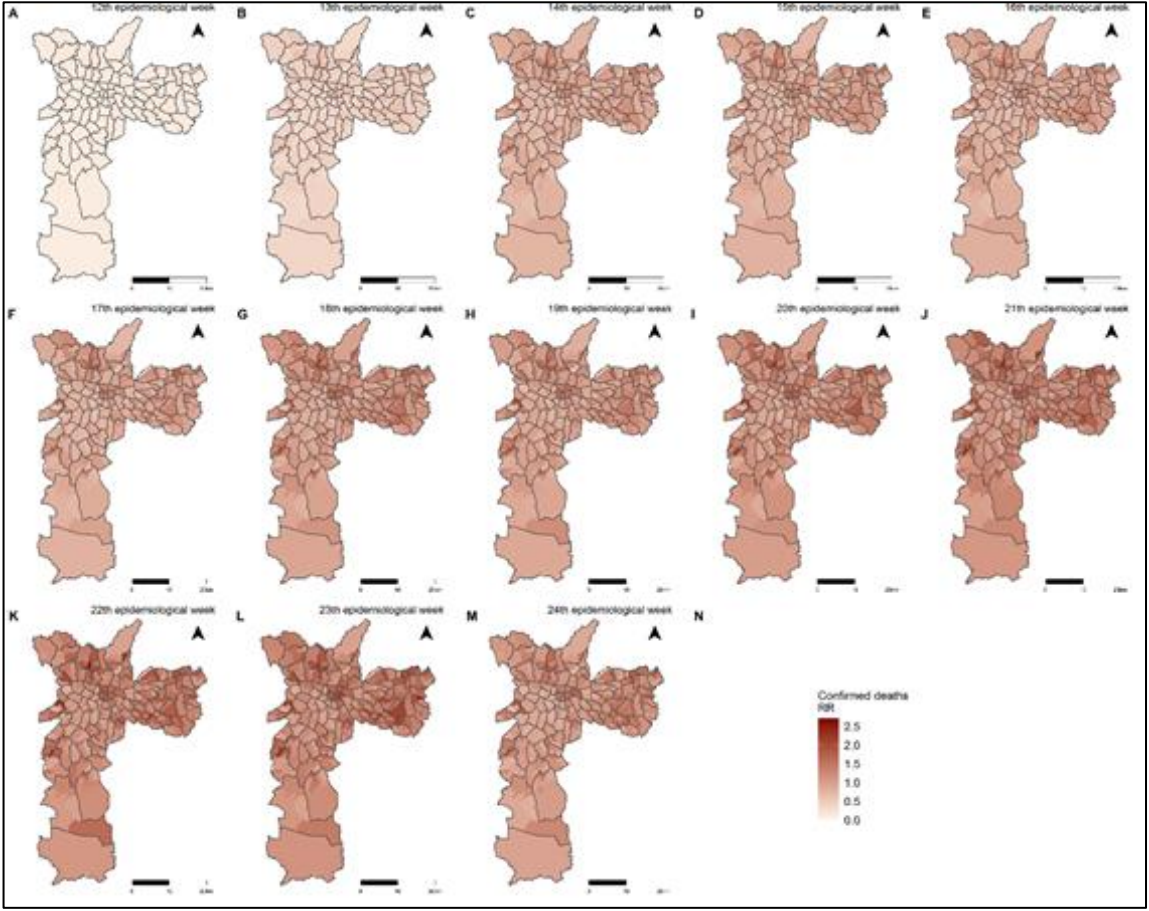

**Figure 5.** Posterior means of the spatiotemporal relative risks (RR) for total COVID-19 deaths by sample areas of the city of São Paulo, 12th to 24th Epidemiological Week, 2020.

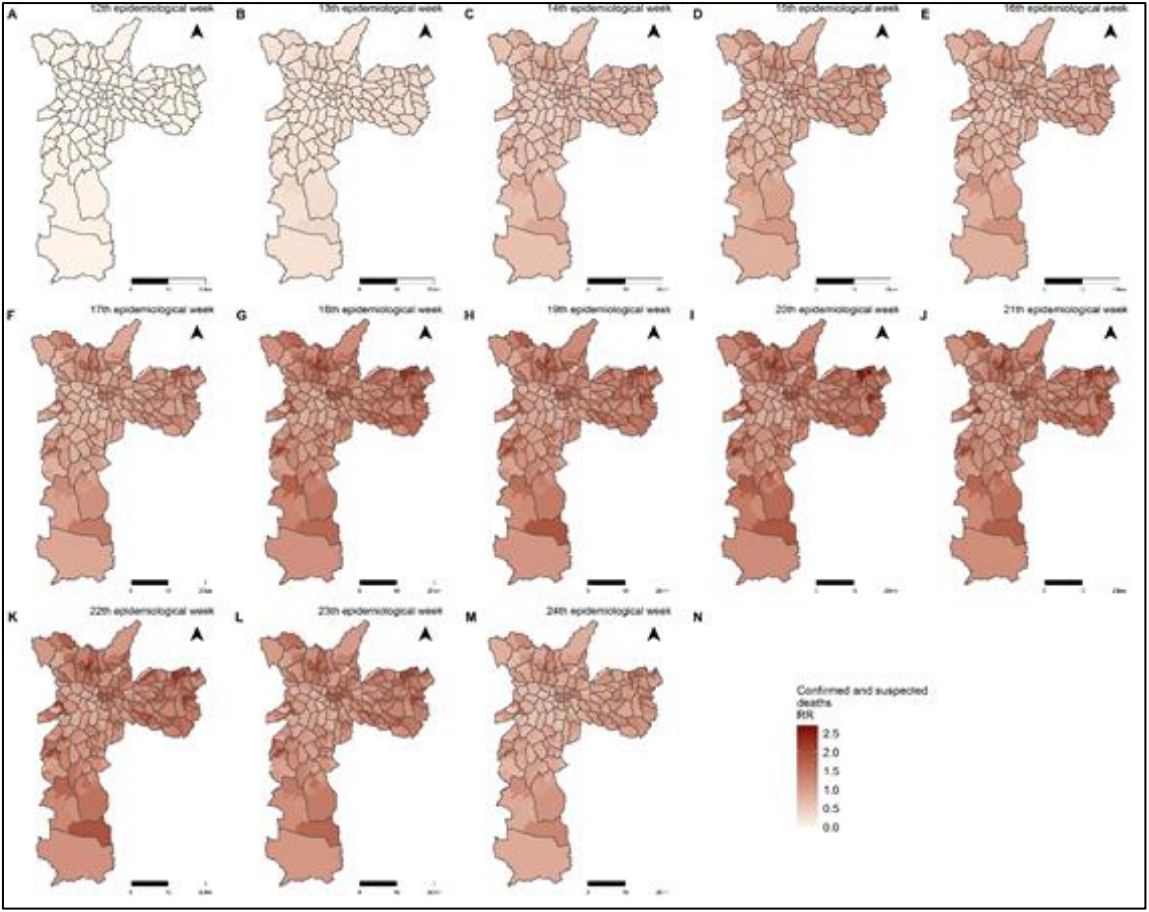

**Figure 6.** Posterior means of the relative risks and 95% credible interval for the socio-economic covariate obtained with spatial models for confirmed and total COVID-19 deaths, according to each one of the epidemiologic weeks. City of São Paulo, 12th to 24th Epidemiology Week, 2020.

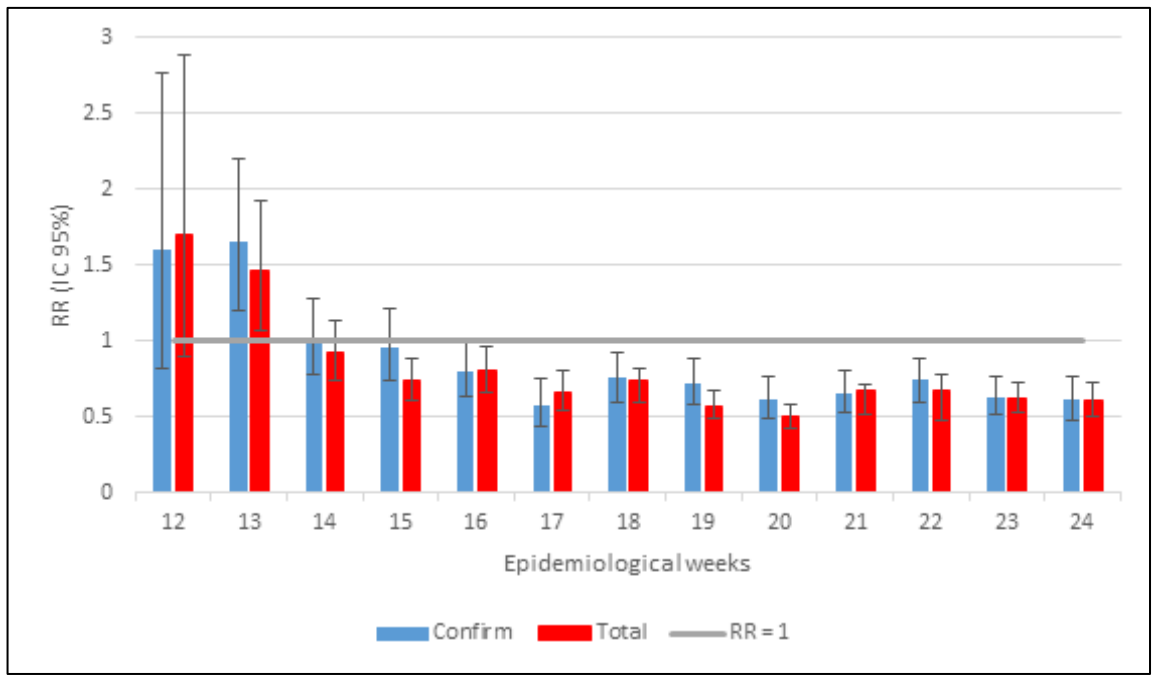

**Supplementary materials**

**Supplementary material 1.** Number and mortality rates (per 100,000 inhabitants in fourteen-weeks) of suspected (U04.9), confirmed (B34.2) and total (U04.9 + B34.2) COVID-19 deaths, according to e-SIC and Tabnet databases and the age groups among people over 60 years of age. City of São Paulo, 11th to 24th epidemiological weeks, 2020.

|  | e-SIC database (provided on 6/18/2020) | | Tabnet database (updated on 7/23/2020) | | Ratio: Tabnet-DataSUS / Database Provided |
|---|---|---|---|---|---|
|  | Nº of deaths | Mortality rate (per 100,000 inhab.) | Nº of deaths | Mortality rate (per 100,000 inhab.) |  |
|  | Total: 60 to 64 years old | | | | |
| **Confirmed** | 593 | 99.8 | 679 | 114.3 | 1.1 |
| **Suspect** | 397 | 66.8 | 348 | 58.6 | 0.9 |
| **Total** | 990 | 166.6 | 1027 | 172.9 | 1.0 |

|  | Total: 65 to 69 years old | | | | |
|---|---|---|---|---|---|
| Confirmed | 667 | 142.4 | 776 | 165.6 | 1.2 |
| Suspect | 499 | 106.5 | 437 | 93.3 | 0.9 |
| Total | 1176 | 251.0 | 1213 | 258.9 | 1.0 |
|  | Total: 70 to 75 years old | | | | |
| Confirmed | 755 | 221.5 | 887 | 260.2 | 1.2 |
| Suspect | 537 | 157.5 | 454 | 133.2 | 0.8 |
| Total | 1292 | 379.0 | 1341 | 393.4 | 1.0 |
|  | Total: 75 years old or older | | | | |
| Confirmed | 2450 | 544.7 | 2836 | 630.5 | 1.2 |
| Suspect | 2306 | 512.7 | 2128 | 473.1 | 0.9 |
| Total | 4756 | 1057.4 | 4964 | 1103.6 | 1.0 |

Data source: Deaths: Sistema de Informações sobre Mortalidade – SIM/PRO-AIM – CEInfo –SMS-SP. Population: Fundação SEADE. Tabnet database was update on 7/23/2020 and e-SIC database was provided on 6/08/2020.

**Supplementary material 2.** DIC values of spatio-temporal models with Poisson and zero-inflated Poisson probability distributions for confirmed and total COVID-19 deaths.

| Considered deaths | Model type | Probability distribution | |
|---|---|---|---|
| | | Poisson | Zero inflated Poisson |
| Confirmed | Intercept | 11880.2 | 11888.7 |
| | Intercept + covariate | 11873.1 | 11887.5 |
| Total | Intercept | 14352.9 | 14354.4 |
| | Intercept + covariate | 14329.6 | 14334.1 |

**Supplementary material 3.** Percentages of sample areas with zero confirmed deaths and zero total deaths by COVID-19 and their respective DIC values for the spatial models with Poisson and zero-inflated Poisson probability distribution, according to epidemiological week (EW)

| EW | Considered deaths | % zeros | Probability distribution and model type | | | |
|---|---|---|---|---|---|---|
| | | | Poisson | | Zero inflated Poisson | |
| | | | intercept | covariate | intercept | covariate |
| 12 | confirmed | 85.5 | 285.7 | 285.9 | 286.4 | 286.7 |
| | total | 84.8 | 305.0 | 304.4 | 304.6 | 304.0 |
| 13 | confirmed | 54.5 | 684.1 | 677.4 | 683.1 | 677.4 |
| | total | 49.7 | 730.0 | 726.4 | 729.2 | 726.9 |
| 14 | confirmed | 31.6 | 928.9 | 930.5 | 928.9 | 930.8 |
| | total | 20.6 | 1079.4 | 1081.0 | 1079.1 | 1080.9 |
| 15 | confirmed | 29.7 | 951.2 | 952.7 | 950.9 | 952.1 |
| | total | 12.6 | 1162.6 | 1155.2 | 1162.9 | 1154.6 |
| 16 | confirmed | 25.8 | 932.1 | 930.2 | 932.5 | 930.4 |
| | total | 11.3 | 1167.3 | 1164.5 | 1167.9 | 1165.3 |
| 17 | confirmed | 29.0 | 950.1 | 941.8 | 952.9 | 943.7 |
| | total | 6.1 | 1214.2 | 1204.9 | 1215.8 | 1205.8 |
| 18 | confirmed | 22.6 | 1023.3 | 1019.5 | 1023.6 | 1019.6 |
| | total | 5.8 | 1321.2 | 1310.4 | 1321.1 | 1309.4 |
| 19 | confirmed | 24.5 | 988.5 | 981.5 | 989.4 | 982.6 |
| | total | 5.1 | 1306.6 | 1275.5 | 1307.9 | 1274.0 |
| 20 | confirmed | 18.7 | 1051.7 | 1035.5 | 1051.4 | 1035.4 |
| | total | 4.2 | 1344.5 | 1289.1 | 1345.8 | 1288.5 |
| 21 | confirmed | 19.0 | 1073.5 | 1063.2 | 1073.5 | 1062.2 |
| | total | 6.5 | 1328.1 | 1306.5 | 1331.2 | 1307.7 |
| 22 | confirmed | 16.1 | 1060.2 | 1052.8 | 1060.6 | 1054.3 |
| | total | 3.5 | 1291.6 | 1267.6 | 1291.9 | 1268.4 |
| 23 | confirmed | 16.4 | 1080.1 | 1060.8 | 1077.9 | 1060.8 |
| | total | 6.1 | 1284.1 | 1257.5 | 1282.7 | 1257.6 |
| 24 | confirmed | 23.8 | 1036.3 | 1028.5 | 1039.8 | 1028.8 |
| | total | 10.0 | 1203.1 | 1187.1 | 1206.9 | 1187.4 |